\documentclass[epj]{svjour}
\usepackage{graphics}
\usepackage{flushend}
\begin{document}
\title{Moment Analysis and Zipf Law}
\author{Y.G. Ma 
}                     
\offprints{E-mail: ygma@sinap.ac.cn 
}          
\institute{Shanghai Institute of Applied Physics, Chinese Academy
of Sciences, Shanghai 201800, China}
\date{Received: date / Revised version: date}
%
\abstract{ The moment analysis method and nuclear Zipf's law of
fragment size distributions are reviewed to study
nuclear disassembly. In this report, we present a compilation of
both theoretical and experimental studies on moment analysis and
Zipf law performed so far. The relationship of both methods to a possible
critical behavior or phase transition of nuclear disassembly is
discussed. In addition, scaled factorial moments and
intermittency are reviewed.
}%
\maketitle
\section{Introduction}
\label{intro}

Hot nuclei can be formed in energetic heavy ion collisions (HIC)
and  deexcite by different decay modes, such as evaporation and
multifragmentation. Experimentally, multifragment
emission was observed to evolve with excitation energy.
The multiplicity, $N_{imf}$, of intermediate mass fragment (IMF) rises
with the beam energy, reaches a maximum, and finally falls to a
lower value. The onset of multifragmentation may indicate
the coexistence of liquid and gas phases \cite{Maprc95}.
Phenomenologically, the mass (charge) distribution of IMF
distribution can be expressed as a power law with parameter
$\tau_{eff}$, and a minimum $\tau_{min}$ of   $\tau_{eff}$
emerges around the onset point, which suggests that a kind of
critical behavior may take place. In the framework of Fisher's
droplet model, the mass distribution can be described by a power
law with a critical exponent of $\tau \sim 2.3$ when the system is
in the vicinity of the critical point \cite{Fisher}.

On the other hand, the caloric curve measurement can also provide
useful information on the liquid-gas phase transition
\cite{Albergo,Poch,Ma-PLB97,JBN0,JBN1}. The analysis of other independent
critical exponents provides additional indications of  critical behavior
of finite nuclear systems \cite{Bau0,Gilk,Ago,Ell1,Ell2,Bau1}. In
addition, more observables have been proposed to sign the liquid-gas
phase transition or critical behavior of nuclei
\cite{Ma2005,Richert_PhysRep,Dasgupta_01,Bonasera,Chomaz_INPC,Moretto_INPC}.
Some reviews can be found in this volume
\cite{JBN-WCI,Viola-WCI,Borderie-WCI,Gul-WCI,Lopez-WCI}.

In this report, we shall review the moment analysis method and
Zipf law of fragment size distribution. The phenomenological basis
of moment analysis is introduced in Sec. 2. Finite size
effects are discussed in Sec.3. Section 4 gives the application of
moment analysis to
 multifragmention and its relation to critical behavior. Scaled
factorial moments and intermittency are discussed in Sec. 5.
 In Sec. 6 Zipf law is introduced for the nuclear fragment
 distribution and the corresponding simulations are given;
some experimental indications of nuclear Zipf law  are
presented in Sec. 7;
 finally the summary and outlook are  given in Sec. 8.

\section{Phenomenological Basis of Moment Analysis}

Campi \cite{Campi-JPA86,Campi-PLB88} and Bauer
\cite{Bauer-PLB85,Bauer-NPA86} {\it{et\ al.}} first suggested that
the methods used in percolation studies may be applied to nuclear
multifragmentation data. In percolation theory the moments of the
cluster distribution contain a signature of critical behavior
\cite{Stauffer}. The method of moment analysis has been
experimentally used to search for evidence of the critical
behavior in multifragmentation. The definition of the $k$ moments
of the cluster size distribution for each event is
\begin{equation}
M_k  = \sum_{A \neq A_{max}} A^k n_A,
\end{equation}
where $A$ is the fragment mass, 
and $n_A$ is the number of charged fragments
whose charge is $Z$ and mass is $A$. The sum runs over all masses
$A$ in the event including neutrons except the heaviest
fragment ($A_{max}$). This quantity was taken as a basic tool in
extracting  critical exponents in Au + C data \cite{Gilk}. It
has been argued that there should be an enhancement in the
critical region of the moment $M_k$, for $k
> \tau -1$, with a critical exponent $\tau > 2$
\cite{Campi-JPA86,Campi-PLB88}.

In experimental analyses, events are sorted by different
conditions. In this case, so-called conditional moments are used
to describe the fragment distribution. Usually the mean value of
$M_k(m)$ for events with given control parameters, e.g. the moment
$M_k$ for events with a given multiplicity $m$, or total bound
charge number $Z_{bound}$, or excitation energy $E^*$, is called
conditional moment.

More insight in the shape of the fragment size distribution is
obtained by looking at a combination of moments $M_k$. For
example, the quantity
\begin{equation}
\gamma_2 = \frac{M_2 M_0}{M_1^2} = \frac{\sigma^2}{\langle s
\rangle^2} + 1,
\end{equation}
has been used,
where $M_1$ and $M_2$ are the first and second moments of the mass
distribution  and $M_0$ is the total multiplicity including
neutrons.  $\sigma^2$ is the variance of the fragment distribution and
$\langle s \rangle = M_1/M_0$ represents the mean fragment size.
$\gamma_2$ takes the value $\gamma_2 = 2$ for a pure exponential
distribution $N(s)\sim exp(-\alpha s)$ regardless of the value of
$\alpha$, but $\gamma_2 \gg 2$ for a power-law distribution $N(s)
\sim s^{-\tau}$ when $\tau > 2$. In the percolation model, the
position of the maximum $\gamma_2$ value defines the critical
point, where the fluctuations in the fragment size distribution are
the largest. In principle a genuine critical behavior requires  the peak
value of $\gamma_2$ to be larger than 2
\cite{Campi-JPA86,Campi-PLB88}. However, due to finite size effects,
this is not always true when the system size decreases, as
we will see in the following sections.

Campi also suggested to use the single event (j) moment, i.e.
\begin{equation}
M_k^{(j)} = \sum_{A \neq A_{max}} A^k n^{(j)}
\end{equation}
to investigate the shape of fragment size distribution. Also
normalized moments \cite{Campi-JPA86}
\begin{equation}
S_k^{(j)} = {M_k^{(j)}}/{M_1^{(j)}}
\end{equation}
can be defined. It was suggested to use the event-by-event scatter-plots
of the natural log of the size ($A_{max}$) or charge number
 ($Z_{max}$) of the largest cluster, ln$A_{max}$ or ln$Z_{max}$ versus the
natural log of the second moment, ln$M_2$, or the normalized
moment ln$S_2$ to search for the largest fluctuation point. Some
examples will be given in the following sections.

In the percolation model, the cluster size distribution for infinite
systems near a critical point  can be expressed by
\begin{equation}
n(s) \sim s^{-\tau} f(\epsilon s^\sigma). \label{ns-scalin}
\end{equation}
where $s$ is the size of finite clusters, $\tau$ and $\sigma$ two
critical exponents and $\epsilon$ a variable that characterizes
the state of the system. In thermal phase transitions $\epsilon =
T - T_c$ is the distance to the critical temperature $T_c$. In
percolation $\epsilon = p_c - p$ is the distance to the critical
fraction of active bonds or occupied sites $p_c$. The scaling
function $f(\epsilon s^\sigma)$ satisfies $f(0)$ = 1, decaying
rapidly (exponentially) for large values of $|\epsilon|$. In
addition, theory predicts that when $\epsilon < 0$ one
infinite cluster (liquid or gel) is present in the system while no
such cluster exists when $\epsilon > 0$ (only droplets or
$n$-mers). In finite systems a similar  behavior is observed,
especially when the largest cluster is counted separately.

The moment analysis method is useful to obtain some information about
the possible occurrence of a critical behavior. In general, critical exponents can be
defined according to the standard procedure followed in condensed
matter physics \cite{Binder}. For example,
\begin{equation}
M_k(\epsilon) = \sum_A = A^k n_A(\epsilon) \sim
|\epsilon|^\frac{\tau-k-1}{\sigma},  (\epsilon \to 0)
\end{equation}
where $\tau$ and $\sigma$ are the critical exponents. For the
 percolation phase transition and the critical point in the
Fisher droplet model, the exponent $\tau$ satisfies $2 < \tau < 3$
and thus the second and high moments diverge at the critical
point. In contrast the lower moments $M_0$ and $M_1$, which
correspond to the number of fragments and the total mass, do not
diverge.

Based upon the scaling relation Eq.~(\ref{ns-scalin}), there
exists the following relationship between critical exponents and
moments:
\begin{eqnarray}
M_0 \sim |\epsilon|^{2-\alpha} \nonumber , \\
M_1 \sim |\epsilon|^{\beta} \nonumber , \\
M_2 \sim |\epsilon|^{-\gamma},
\end{eqnarray}
where $\beta$ and $\gamma$ are two other critical exponents. Some
relationships among critical exponents exist (hyperscaling relations), for instance
\begin{equation}
2\beta + \gamma = \frac{\tau-1}{\sigma} = 2 - \alpha .
\end{equation}

In finite systems transitions are smooth, but it is still
possible to determine some critical exponents, as we will discuss in the next section.
By analogy with
the infinite system behavior, one says that these moments exhibit a
critical behavior also for finite systems. In particular in the
Fisher model, the thermal critical point is also
a critical point for moments of the fragment size distribution.

In order to illustrate the application of moment analysis, we show
the EOS data and NIMROD data as examples in Sec. 4.

\section{Finite Size Effects}

Since the nucleus is a finite size system, the macroscopic thermal
limit cannot be applied. Therefore finite size effects on
phase transition behavior should be checked. In this section, we
give some examples to illustrate this problem.

A percolation on a cubic lattice of linear size $L$ containing
$L^3$ sites, for $L$ = 4 to 10, where all  sites are occupied
and bonds are assumed to exist between neighbouring sites with
bond probability $p$, has been considered \cite{Jaqaman}. Sites
that are connected together by such bonds are said to belong to
the same cluster. It is well known that in such a model there
exists a critical (or threshold) probability $p_c$ such that for
$p
> p_c$ there is a large cluster that percolates throughout the
lattice from end to end whereas for $p < p_c$ no such cluster
exists and all the sites belong to small clusters (including
isolated sites, i.e. singlet or clusters of size 1). As $L
\to \infty$ the transition becomes sharper and $p_c$
approaches a limiting value which for bond percolation on a cubic
lattice is $p_c$ = 0.249 \cite{Stauffer}. For finite systems the
threshold percolation probability is not so sharply defined.

\begin{figure}
\begin{center}
\resizebox{1.0\columnwidth}{!}{%
 \includegraphics{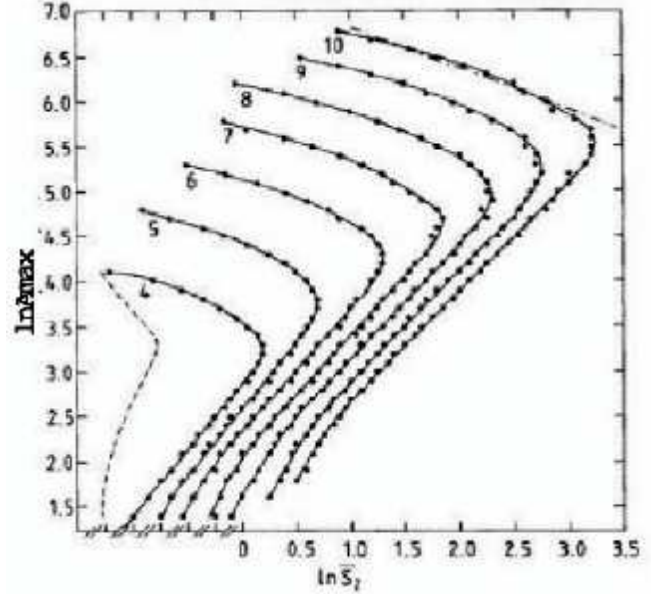}  }
\end{center}
\caption{  The logarithm of the largest fragment
  size $A_{max}$ as a function of the logarithm of the corresponding average normalized second moment
  $\overline{S_2}$ for bond percolation on simple cubic lattices of linear size ranging
  from $L$ = 4 - 10 sites. The dots represent the actual calculation results
  and the curves drawn are just to guide the eye. The  number next to each curve gives the value of the linear size
  $L$. Note that the ln$\overline{S_2}$ scale given corresponds to the $L$ = 10
  curve. The other curves are successively shifted to the left with
  respect to each other by a distance of 0.25. The dashed curve and the
  dotted-dashed straight line are explained in the text. Figure is taken from Ref.~\cite{Jaqaman}.}
  \label{fig_Campi_size}
\end{figure}

In order to quantitatively illustrate finite size effects on
critical behavior, the average of normalized second moment
$S_2$ ($\overline{S_2}$) over all events belonging to the same
value of ln($A_{max}$) was calculated \cite{Jaqaman}. The results
obtained by such averaging are presented by the dots shown in
Fig.~\ref{fig_Campi_size} for various cubic lattices with linear
dimension $L$ = 4 - 10 sites \cite{Jaqaman}. The location of the
maximum value of $\overline{S_2}$ is now defined as corresponding
to the location of the critical point, which is a standard way of
determining the percolation threshold \cite{Jan}. The slope of the
lower branches of the curves in Fig.~\ref{fig_Campi_size} can also
be calculated. This slope is expected to be 1 + $\beta/\gamma$
which for percolation in three dimensions is equal to 1.23. For
comparison the slopes of the straight lines by a lest-squares fit
to the lower branches of the $L$ = 4 to 10 curves are found, in
ascending order of $L$, to have the values 1.582$\pm$0.036,
1.503$\pm$0.029, 1.375$\pm$0.017, 1.355$\pm$0.021,
1.260$\pm$0.007, 1.258$\pm$0.014 and 1.242$\pm$0.015
\cite{Jaqaman}. This indicates that these slopes rapidly approach
the value expected in the thermodynamic limit. In calculating
these slopes one has excluded the points near the bottom of the
branch in the region where the curves in Fig.~\ref{fig_Campi_size}
deviate noticeably from a straight line. These points correspond
to events that are far from the critical region.

\begin{figure}
\begin{center}
\resizebox{0.74\columnwidth}{!}{%
  \includegraphics{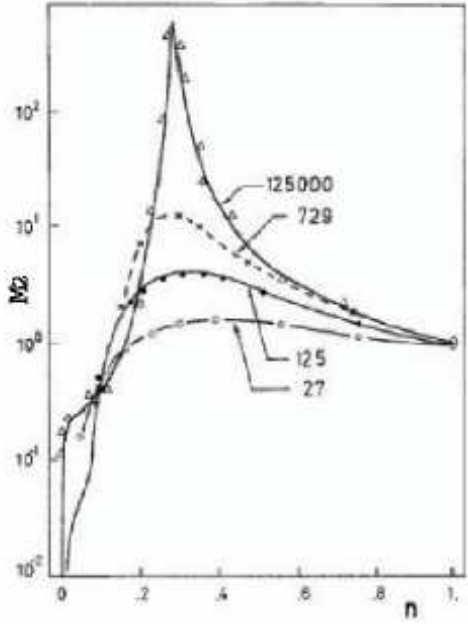}}
\end{center}
\caption{The conditional moments $M_2(n)$ for percolation in a
cubic lattice of linear size $L$ = 3, 5, 9 and 50 (the
corresponding cubic lattice is $L^3$ which is shown as the number
in the insert). The figure is taken from Ref.~\cite{Campi-PLB88}.
} \label{fig_Campi_fig2}
\end{figure}

Similarly to the analysis for the correlation of $\overline{S_2}$
and  ln($A_{max}$), finite size effects have been also
investigated for $M_2$ by Campi \cite{Campi-PLB88}. This is shown
in Fig.~\ref{fig_Campi_fig2}, where $M_2(n)$ is plotted for
various system sizes ($50^3, 9^3, 5^3$ and $3^3$) in a percolation
model. We see clearly the critical behavior for the largest
system, namely a well defined peak, and how this peak is smoothed
when decreasing the size \cite{Campi-PLB88}.

\section{Application of the Moment analysis method}

\subsection{EOS data}

\subsubsection{Experimental description}

The reverse kinematic EOS experiment was performed with 1 A GeV
$^{197}$Au, $^{139}$La, and $^{84}$Kr beams on carbon targets. The
experiment was done with the EOS Time Project Chamber (TPC) and
multiple sampling ionization chamber (MUSIC II). The excellent
charge resolution of this detector permitted identification of all
detected fragments. The fully reconstructed multifragmentation
events for which the total charge of the system was taken as $79
\leq Z \leq 83$, $54 \leq Z \leq 60$, $33 \leq Z \leq 39$ for Au,
La, and Kr, respectively
\cite{SrivastavaPRC65,Scharenberg,HaugerPRC57,HaugerPRC62} were
analyzed. The remnant refers to the equilibrated nucleus formed
after the emission of prompt particles. The charge and mass of the
remnant were obtained by removing for each event the total charge
of the prompt particles. The excitation energy of the remnant
$E^*$ was based on an energy balance between the excited remnant
and the final stage of the fragments for each event \cite{Cussol}.
The thermal excitation energy $E_{th}^*$ of the remnant was
obtained as the difference between $E^*$ and $E_x$ which is a
nonthermal component, namely an expansion energy
\cite{SrivastavaPRC65,Scharenberg,HaugerPRC57,HaugerPRC62,LauretPRC57}.

\subsubsection{Determination of Critical Point and Exponent  in Terms of Moment Analysis}

The determination of the critical point and the associated
exponents in the multifragmentation of gold nuclei was first
attempted by the EOS collaboration \cite{Gilk}. In their early
publication \cite{Gilk}, they use the multiplicity $m$, as a
control variable for the collision violence and assume that $m$ is
a linear measure of the distance from the critical point. Then the
critical exponents $\beta$, $\gamma$ and $\tau$,  can be
determined according to Eqs.~(7,8) above. They find that these
exponents are close to the nominal liquid-gas universality class
values. However, this method is very delicate. In particular, due
to the small size of the system, an important rounding of the
transition is expected which may distort considerably the
determined critical exponents. For a review of this debate, see
the arguments between Bauer \cite{Bauer2} and Gilkes \cite{Gilk2}.

A different analysis was also proposed
by the EOS collaboration \cite{Elliott_PRC03}.
In this work, thermal excitation energy has
been taken as  a control variable, which is believed to be more
suitable to characterize the collision violence.

\begin{figure}
\begin{center}
\resizebox{0.73\columnwidth}{!}{%
  \includegraphics{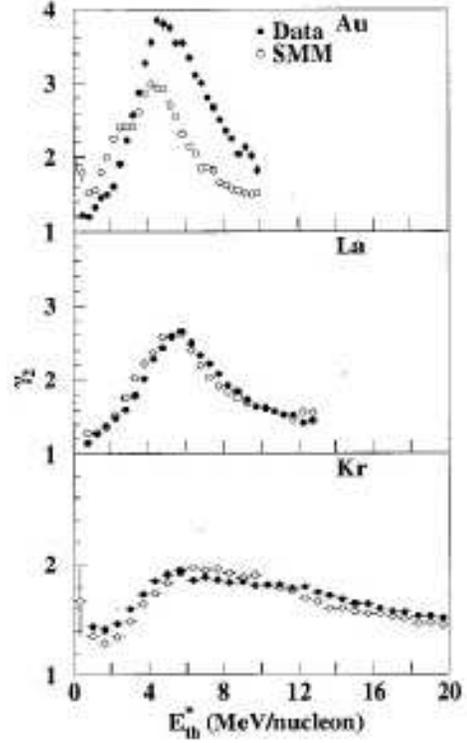}}
\end{center}
\caption{$\gamma_2$ as a
function of $E_{th}^*$  for all three systems of 1 A GeV Au, La,
and Kr collisions with C target and SMM calculations. Figure is
taken from Ref.~\cite{SrivastavaPRC65}.} \label{EOS_gamma2}
\end{figure}

The $\gamma_2$ analysis is shown in Fig.~\ref{EOS_gamma2} for all
three systems. The position of the maximum $\gamma_2$ value
defines the critical excitation energy $E^*_c$, which corresponds
to the largest fluctuation point in the fragment size distribution.
The peak in $\gamma_2$ is well defined for La and Au. For Kr, the
peak is very broad and the value $\gamma_2$ is less than 2.

Fig.~\ref{EOS_gamma2} also shows a $\gamma_2$ calculation using
the statistical multifragmentation model (SMM).
The fission contribution to $\gamma_2$ has been removed
both from the data and SMM.
In the case of Au, the
$\gamma_2$ value remains above two for most of the excitation
energy range both in data and SMM.
The $E_{th}^*$ width over which $\gamma_2 > 2$ is
smaller for La and disappears  for Kr. The decrease in $\gamma_2$
with decreasing  system size is also seen in 3D percolation
studies and these differences have been attributed to finite size
effects \cite{Elliott_PRC03,Campi92a,Campi92b}.

The exponent $\tau$ can be obtained if the second moment $M_2$ and
the third moment $M_3$ of the fragment mass distributions are known. A
plot of ln($M_3$) vs ln($M_2$) should give a straight line with a
slope given by
\begin{equation}
S = \frac{\Delta ln(M_3)}{\Delta ln(M_2)} = \frac{\tau -4}{\tau
-3}.
\label{eq_tau}
\end{equation}

\begin{figure}
\begin{center}
\resizebox{1.0\columnwidth}{!}{%
  \includegraphics{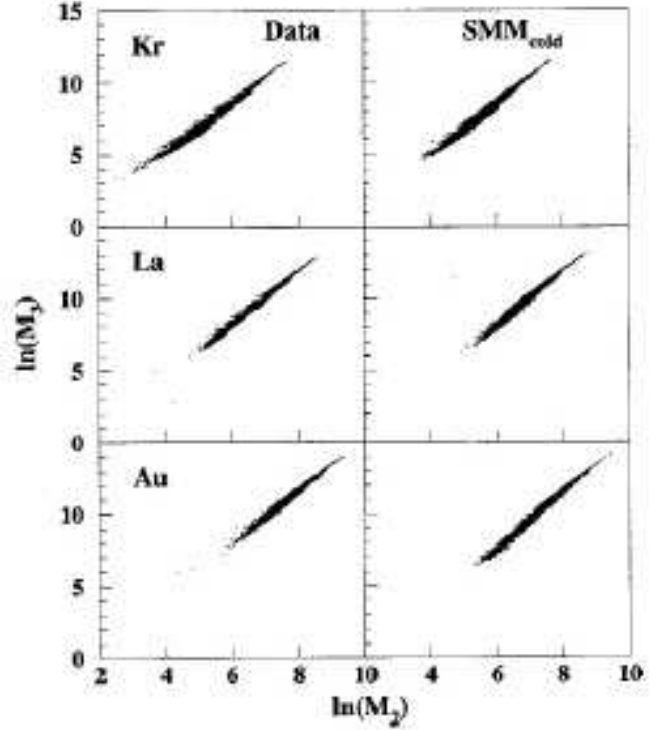}  }
\end{center}
\caption{ln($M_3$) vs ln($M_2$)
for Au, La, and Kr above the critical energy. Figure is taken from
Ref.~\cite{SrivastavaPRC65}.} \label{EOS_M32}
\end{figure}

Fig.~\ref{EOS_M32} shows a scatter-plot of ln($M_3$) vs ln($M_2$)
for the three systems constructed with data above the critical excitation
energy $E_c^*$ (see Fig.~\ref{EOS_gamma2}) and with SMM
simulations. A linear fit to ln($M_3$) vs ln($M_2$) gives the
value of $\tau$. The fitted $\tau$ values are $2.16 \pm 0.08$, $2.10
\pm 0.06$ and $1.88 \pm 0.08$, respectively. The former two are
very close to the critical exponents $\tau \sim 2.3$ of the liquid-gas
universal class.

\begin{figure}
\begin{center}
\resizebox{.8\columnwidth}{!}{%
  \includegraphics{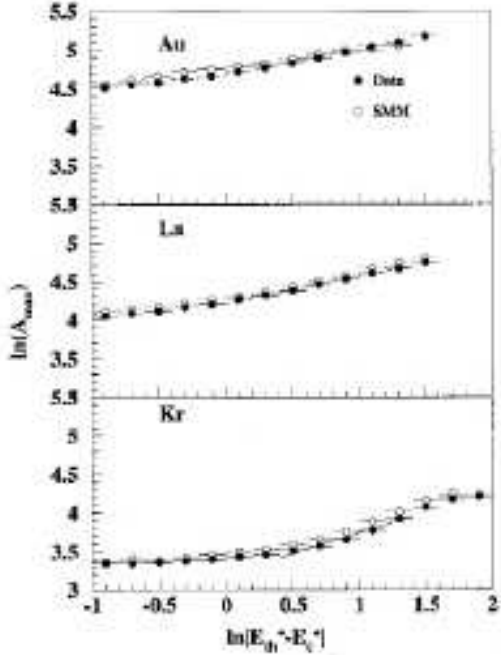}  }
\end{center}
\caption{ln($A_{max}$) vs ln$|E^*_{th}-E_C^*|$ for Au,
La, and Kr below the critical energy for exponent $\beta$
determination. Figure is taken from Ref.~\cite{SrivastavaPRC65}.}
\label{fig_EOS_gamma}
\end{figure}

The exponent $\beta$ can be obtained for the multifragmentation
data by the relation
\begin{equation}
A_{max} \sim |\epsilon|^\beta,
\end{equation}
where $\epsilon = p - p_c$ and $\epsilon > 0$. In the
multifragmentation case of this work $p$ and $p_c$ have been
replaced by $E^*_{th}$ and $E_c^*$. In an infinite system, the
finite cluster exists only on the liquid side of $p_c$. In a
finite system a largest cluster is present on both sides of the
critical point, but the above equation holds only on the liquid
side. Fig.~\ref{fig_EOS_gamma} shows a plot of ln$(A_{max})$ vs
ln$|E^*_{th}-E_c^*|$ for Au, La, and Kr. The values of $\beta$
extracted for Au and La are 0.32$\pm$0.02 and 0.34$\pm$0.02,
respectively, which are close to the value of 0.33 predicted for a
liquid-gas phase transition. On the other hand, the value of
$\beta$ = 0.53 $ \pm$ 0.05 for Kr is much higher than that of Au
and La.

\begin{figure}
\begin{center}
\resizebox{1.0\columnwidth}{!}{%
  \includegraphics{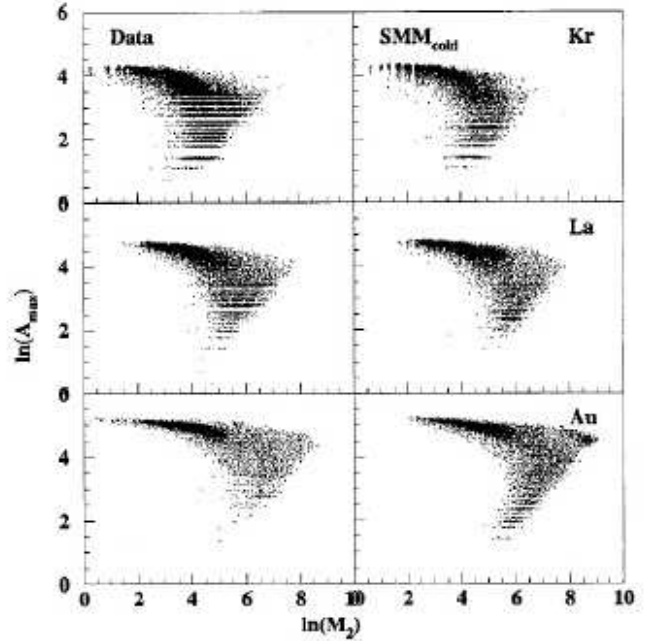}}
\end{center}
\caption{Scatter-plots of ln($A_{max}$)   vs ln($M_2$) from the
data for Au, La, and Kr. Left panel: EOS data; Right panel: SMM
simulation. The figure is taken from Ref.~\cite{SrivastavaPRC65}.
} \label{fig_Campi_plot}
\end{figure}

As shown in Sec. 2, Campi also suggested that the correlation
between the size of the biggest fragment $A_{max}$ and the moments
in each event, i.e. the scatter-plot, can measure the critical
behavior in nuclei. Fig.~\ref{fig_Campi_plot} depicts a scatter-
plot with logarithmic scale for Au, La, and Kr of EOS data.  The
two branches corresponding to the sub-critical (upper branch) and
overcritical (lower branch) events are clearly seen for Au and La.
The scatter-plot is very broad for Kr and fills most of the
available space. The sub- and over-critical branches seem to
overlap and are not well separated. Studies on  SMM show a
similar behavior. If one knows the location of the
critical point from some other
methods, then the scatter-plot can be used to calculate the ratio
of critical exponents $\beta/\gamma$ from the slope of the
sub-critical branch. In EOS data, the position of the largest
$\gamma_2$ was used to define the critical point, which
corresponds to the largest fluctuation of the fragment
distribution. In this context, $\beta/\gamma$ values for Au, La
and Kr can be extracted from the linear fit to the upper branch;
they are 0.22 $\pm$ 0.03, 0.25 $\pm$ 0.01 and 0.50 $\pm$ 0.01,
respectively. $\beta/\gamma$   values of Au and La are close to the value
0.26 expected for the liquid-gas universality class.

To summarize the critical exponent analysis of the EOS data, the
experimental results in conjunction with SMM provide some
indications on the order of the phase transition in Au, La and Kr.
The values of the critical exponents $\tau$, $\beta$, and $\gamma$,
which are close to the values  of a liquid-gas system, along with
nearly zero latent heat (this subject  is beyond the discussion
topics in this review, but the interested reader is reported to
Refs.~\cite{SrivastavaPRC65,Scharenberg})
have been interpreted by the authors as
 a continuous phase
transition in Au and La. However, the analysis of Kr leads to very different
critical exponents. A recent analysis based on the shape of
SMM microcanonical caloric curve indicates a first order phase
transition for the multifragmentation of Kr
\cite{SrivastavaPRC65,Scharenberg}.

\subsection{NIMROD data}

\subsubsection{Experimental set-up and Analysis Details}

Using the TAMU NIMROD (Neutron Ion Multidetector for Reaction
Oriented Dynamics) and beams from the TAMU K500 super-conducting
cyclotron, we have probed the properties of excited
projectile-like fragments produced in the reactions of 47
MeV/nucleon $^{40}$Ar + $^{27}$Al, $^{48}$Ti and $^{58}$Ni. The
charged particle detector array of NIMROD, which is set inside a
neutron ball, includes 166 individual CsI detectors arranged in 12
rings in polar angles from $\sim$ $3^\circ$  to $\sim$
$170^\circ$. The detailed description for the experiment  can be
found in \cite{Ma2005}. The correlation of the charged particle
multiplicity ($M_{cp}$) and the neutron multiplicity ($M_n$) was
used to sort the event violence. After the reconstruction of the
quasi-projectile (QP)
particle source, the excitation energy was deduced event-by-event
using the energy balance equation \cite{Cussol}.

\subsubsection{Critical Point Determination via Moment Analysis}

\begin{figure}
\begin{center}
\resizebox{0.9\columnwidth}{!}{%
  \includegraphics{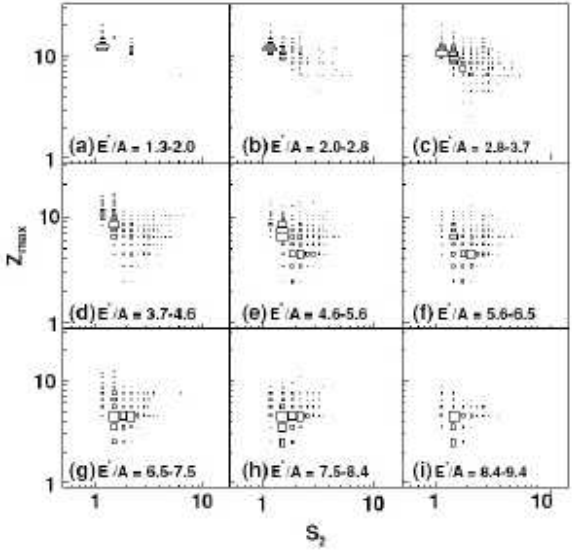}  }
\end{center}
\caption{Campi plots for nine intervals of excitation
energy for the QP formed in $^{40}$Ar + $^{58}$Ni. Figure
is taken from Ref.~\cite{Ma2005}.
 } \label{fig_NIMROD_Campi_plot}
\end{figure}

In Fig.~\ref{fig_NIMROD_Campi_plot} we present Campi scatter-plots for the
nine selected excitation energy bins. In the low excitation energy
bins of $E^*/A$ $\leq$ 3.7 MeV/u,   the upper (liquid phase)
branch is strongly dominant   while at $E^*/A$ $\geq$ 7.5 MeV/u,
the lower $Z_{max}$ (gas phase) branch is strongly dominant. In
the region of intermediate  $E^*/A$   of 4.6- 6.5 MeV/u, the
transition from the liquid dominated branch to the vapor branch
occurs, indicating that the region of maximal fluctuations is to
be found in that range.

\begin{figure}
\begin{center}
\resizebox{.9\columnwidth}{!}{%
  \includegraphics{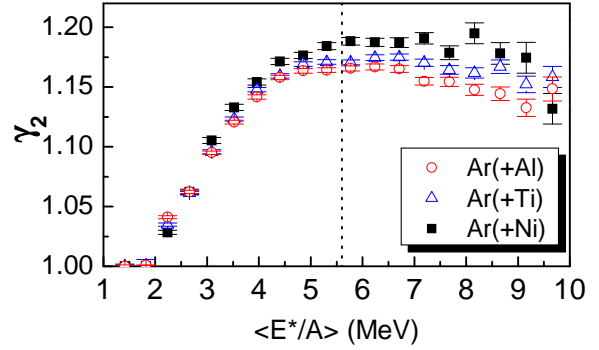}  }
\end{center}
\caption{$\gamma_2$ of the QP
systems formed in Ar + Al (open circles), Ti (open triangles) and
Ni (solid squares) as a function of excitation energy.
 } \label{fig_NIMROD_g2}
\end{figure}

The excitation energy dependence of the average values of
$\gamma_2$  obtained in an event-by-event analysis of our data are
shown in Fig.~\ref{fig_NIMROD_g2}. $\gamma_2$ reaches its maximum
in the 5-6 MeV excitation energy range. In contrast to
observations for heavier systems of Au and La
\cite{SrivastavaPRC65,Elliott_PRC03}, there is no well defined
peak in $\gamma_2$ for our very  light system and $\gamma_2$ is
relatively constant at higher excitation energies. This is similar
to the case of Kr of EOS data.  We note also that the peak value
of  $\gamma_2$ is lower than 2 which is the expected smallest
value for critical behavior in large systems. However, 3D
percolation studies indicate that finite size effects can lead to
a decrease of $\gamma_2$ with system size
\cite{Campi92a,Campi92b}. For a percolation system with 64 sites,
peaks in $\gamma_2$ under two are observed. Therefore, the
criterion $\gamma_2 > 2$  alone is not sufficient to discriminate
whether or not the critical point is reached.

\begin{figure}
\begin{center}
\resizebox{.9\columnwidth}{!}{
\includegraphics{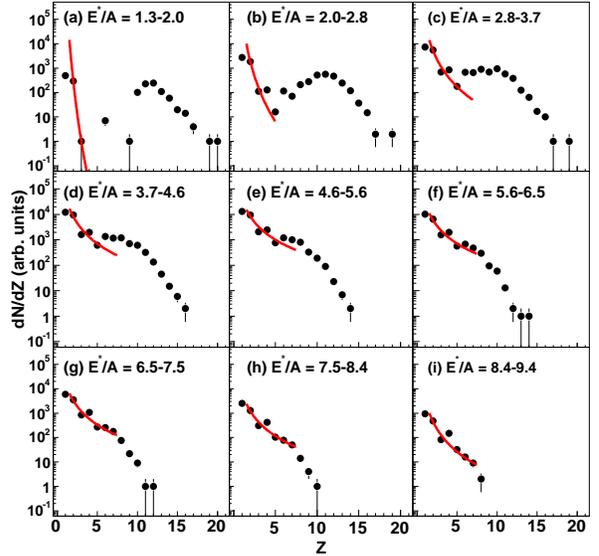}}
\end{center}
\caption{Charge distribution of QP in different
$E^*/A$ window for the reaction  $^{40}$Ar + $^{58}$Ni. Lines
represent fits.  Figure is taken from Ref.~\cite{Ma2005}. }
\label{fig_Zdist}
\end{figure}

\begin{figure}
\begin{center}
\resizebox{0.9\columnwidth}{!}{%
  \includegraphics{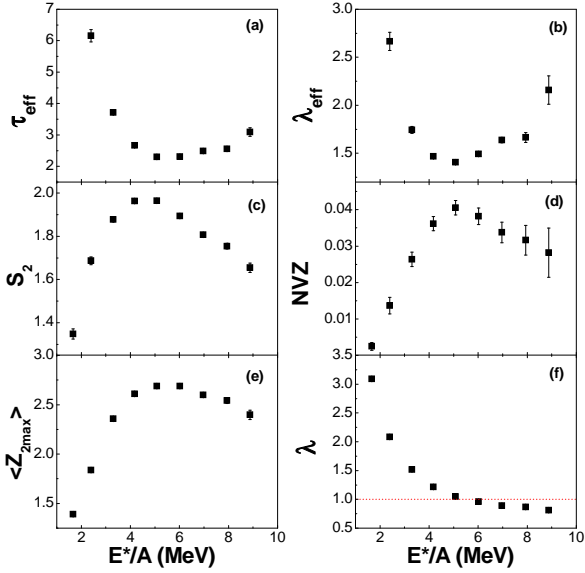}}
\end{center}
\caption{The effective Fisher-law parameter
($\tau_{eff}$) (a), the effective exponential law parameter
($\lambda_{eff}$) (b), $\langle S_2\rangle$ (c), NVZ fluctuation
(d),   the mean charge number of the second largest fragment
$\langle Z_{2max}\rangle$ (e), the Zipf-law parameter $\lambda$
(f). See details in text.  Figure is taken from
Ref.~\cite{Ma_NPA2005}.}
 \label{fluctuation}
\end{figure}

In the Fisher droplet model, the critical exponent $\tau$ can be
deduced from the cluster distribution near the critical  point. To
quantitatively pin down the possible phase transition point, we
use a power law fit to the QP charge distribution in the range of
$Z$ = 2 - 7  (Fig.~\ref{fig_Zdist})
to extract the effective Fisher-law parameter
$\tau_{eff}$ by
\begin{equation}
dN/dZ \sim Z^{-\tau_{eff}}. \label{equ_tau_eff}
\end{equation}
Fig.~\ref{fluctuation}(a) shows the effective Fisher-law parameter
$\tau_{eff}$ as a function of excitation energy. A minimum with
$\tau_{eff}$ $\sim$ 2.3 is seen to occur in the $E^*/A$ range of 5
to 6 MeV/u \cite{Ma_NPA2005}. This value is close to the critical
exponent of the liquid-gas phase transition universality class \cite{Fisher}.

Assuming that the heaviest cluster in each event represents
the liquid phase, we have attempted to isolate the gas phase by
event-by-event removal of the heaviest cluster from the charge
distributions. We find that the resultant distributions are better
described with an exponential form $ exp^{-\lambda_{eff} Z}$. The
fitting parameter $\lambda_{eff}$  was derived and is plotted
against excitation energy in Fig.~\ref{fluctuation}(b). A minimum
is seen in the same region where $\tau_{eff}$ shows a minimum. To
further explore this region we have investigated other proposed
observables commonly related to fluctuations and critical
behavior. Fig.~\ref{fluctuation}(c) shows the mean normalized
second moment, $\langle S_2\rangle$ as a function of excitation
energy. A peak is seen around 5.6 MeV/u, it indicates that the
fluctuation of the fragment distribution is the largest in this
excitation energy region. Similarly,  the normalized variance in
$Z_{max}/Z_{QP}$ distribution (i.e. NVZ =
$\frac{\sigma^2_{Z_{max}/Z_{QP}}}{\langle Z_{max}/Z_{QP}\rangle}$)
\cite{Dorso} shows a maximum in the same excitation energy region
(Fig.~\ref{fluctuation}(d)), which illustrates the maximal
fluctuation for the largest fragment is reached
around $E^*/A$ = 5.6 MeV. The second largest fragment
shows a behavior similar to the
one of the largest fragment.
Fig.~\ref{fluctuation}(e) shows a broad peak of $\langle
Z_{2max}\rangle$ - the average atomic number of the second largest
fragment - also occurring in the same excitation energy range around 5.6 MeV/u.

\begin{figure}
\begin{center}
\resizebox{.85\columnwidth}{!}{
\includegraphics{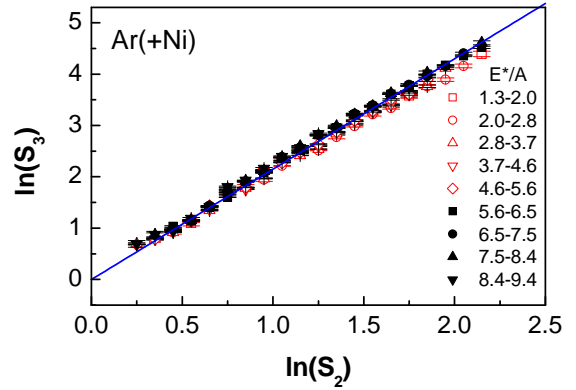}}
\end{center}
\caption{The
correlation between ln($S_3$) vs ln($S_2$) and a linear fit.
} \label{fig_tau_s23}
\end{figure}

 More variables have been collected to support the
determination of the critical point around 5.6 MeV/u of excitation energy for
our system \cite{Ma2005}, such as $\Delta$-scaling \cite{Botet} or
energy fluctuations \cite{Gul-WCI}. In addition, the
measurement of the caloric curve \cite{Ma2005} gives a temperature,
$T_c \sim$ 8.3 MeV around $E^*/A$ = 5.6 MeV.
The value of the critical temperature is needed for
the determination of the critical exponents, as explained in the following
subsection.

\subsubsection{Determination of Critical Exponents Based on Moment Analysis }

In terms of the scaling theory, $\tau$ can also be deduced from
 Eq.~(\ref{eq_tau}). 
Since the value of $T_c$ = 8.3 MeV has been determined from our
caloric curve measurements \cite{Ma2005}, we can explore the
correlation of $S_2$ and $S_3$ in two ranges of excitation energy
(see Figure~\ref{fig_tau_s23}). The moments were calculated by
excluding the species with $Z_{max}$  for the "liquid" phase but
including it in the "vapor" phase. The slopes were determined from
linear fits to the "vapor" and "liquid" regions respectively and
then averaged. In this way, we obtained a   value of $\tau = 2.13
\pm 0.1$.

 Other  critical exponents
can also be related to other moments of the cluster distribution,
$M_k$. Using our caloric curve measurements \cite{Ma2005},
we can use temperature as a control
parameter for such determinations.  Then the critical
exponent $\beta$ can be extracted from the relation
\begin{equation}
Z_{max} \propto (1-\frac{T}{T_c})^\beta, \label{eq_beta}
\end{equation}
and the critical exponent $\gamma$ can be extracted from the
second moment via
\begin{equation}
M_2 \propto |1-\frac{T}{T_c}|^{-\gamma}. \label{eq_gamma}
\end{equation}
In both equations, $|1-\frac{T}{T_c}|$ is the parameter which measures the
distance from the critical point.

\begin{figure}
\begin{center}
\resizebox{.9\columnwidth}{!}{
\includegraphics{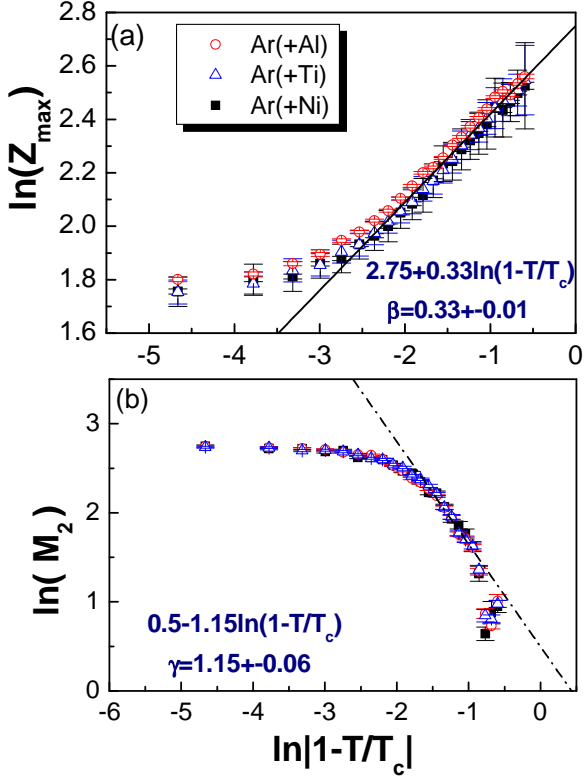}}
\end{center}
\caption{ The extraction of the
critical exponent $\beta$ (a) and $\gamma$ (b). See texts for
details. } \label{fig_beta-gamma}
\end{figure}

The upper panel of Fig.~\ref{fig_beta-gamma}  explores the
dependence of $Z_{max}$ on $(1- \frac{T}{T_c})$. A dramatic change
of $Z_{max}$ around the critical temperature  $T_c$ is observed.
Lattice-gas model (LGM)
calculations also predict that the slope of $Z_{max}$ vs $T$
will change at the liquid-gas phase transition \cite{Ma_JPG01}.
Using the liquid side points, we can deduce the critical
exponent $\beta$ by $ln(Z_{max})$ vs $ln|1-T/T_c|$.
Fig.~\ref{fig_beta-gamma}(a) shows the extraction of $\beta$ using
Eq.~(\ref{eq_beta}).  An excellent fit was obtained in the region
away from the critical point, which indicates a critical exponent
$\beta$ = 0.33 $\pm$ 0.01. Near  the critical point, finite
size effects become stronger so that the scaling law is violated.
The extracted value of $\beta$ is that expected for a liquid-gas
transition (See Table 1) \cite{Stauffer}.

To extract the critical exponent $\gamma$, we take $M_2$ on the
liquid side without $Z_{max}$.
Fig.~\ref{fig_beta-gamma}(b) shows ln($M_2$) as a function of
ln($|1-\frac{T}{T_c}|$). We center our fit to Eq.~(\ref{eq_gamma})
about the center of the range of $(1-T/T_c)$ which leads to the
linear fit and extraction of $\beta$ as represented in
Figure~\ref{fig_beta-gamma}. We obtain a critical exponent
$\gamma$ = 1.15 $\pm$ 0.06. This value of $\gamma$ is also close
to the value expected for the
 liquid-gas universality class (see Table 1).
It is seen that the selected region has a good power law
dependence.

Since we have the critical exponent $\beta$ and $\gamma$, we can
use the scaling relation
\begin{equation}
\sigma = \frac{1}{\beta + \gamma},
\end{equation}
to derive the critical exponent $\sigma$. In such way, we get the
$\sigma$ = 0.68 $\pm$ 0.04, which is also very close to the
expected critical exponent of a liquid-gas system.

To summarize the critical exponents extracted from NIMROD data, we present the
results in Table 1 as well as the values expected for the 3D
percolation and liquid-gas  universality classes. It is apparent that
our values for this light system with A$\sim$36 are closer to
the values of the liquid-gas phase transition universality
class rather than to the 3D percolation class.

\begin{table}
\caption{Comparison of the Critical Exponents} \label{tab:1}
\begin{center}
\begin{tabular}{llll}
\hline\noalign{\smallskip}
Exponents & 3D Percolation & Liquid-Gas & NIMROD \\
\noalign{\smallskip}\hline\noalign{\smallskip}
$\tau$ & 2.18 &  2.21 & 2.13$\pm$0.10 \\ 
$\beta$ & 0.41 & 0.33 & 0.33$\pm$0.01 \\
$\gamma$ & 1.8 & 1.23 & 1.15$\pm$0.06 \\
$\sigma$ & 0.45 & 0.64 & 0.68$\pm$0.04 \\
\noalign{\smallskip}\hline
\end{tabular}
\end{center}
\end{table}

\section{Scaled factorial moments and intermittency}

Intermittency is related to the existence of large
non-statistical fluctuations and is a signal of self-similarity of
the fluctuation distribution at all scales. This signal can
be deduced from the scaled factorial moments \cite{Bialas},
\begin{equation}
F_k(\delta) = \frac{\Sigma_{i=1}^{X_{max}/\delta}\langle
n_i(n_i-1)(n_i-2)...(n_i-k+1)\rangle}{\Sigma_{i=1}^{X_{max}/\delta}
\langle n_i \rangle ^k}
\end{equation}
where $X_{max}$ is an upper characteristic value of the system
(i.e. total mass or charge, maximum transverse energy or momentum,
etc.) and $k$ is the order of the moment. The total interval
0-$X_{max}$ (1-$A_{max}$, $Z_{max}$ in the case of mass or charge
distributions) is divided into $X_{max}/\delta$ bins of the size
$\delta$, $n_i$ is the number of particles in the $i$th bin for an
event, and the ensemble average $\langle \rangle$ is performed
over all events. The concept of intermittency was originally
developed in the field of fluid dynamics to study the fluctuations
occurring in turbulent flows \cite{Mandelbrot,Zeldovich}. Its
presence in the velocity and temperature distributions is
established by the existence of large non-statistical fluctuations
which exhibit scale invariance. Intermittency in physical systems
is studied by examining the scaling properties of the moments of
the distributions of relevant variables over a range of scales
\cite{Paladin}. The concept of intermittency  was first introduced
for the study of dynamical fluctuations in the density
distribution of particles produced in high energy collisions by
Bialas and Peschanski \cite{Bialas}. It soon led to the discovery
of  a characteristic power law dependence of the factorial
moments, $F_k$, of an order $k$ on the resolution scale, $\delta$:
$F_k \propto (1/\delta)^{f(k)}$. The specific properties of the
intermittency exponent, $f(k)$, can be associated either with a
random production process \cite{Bialas,Bialas2} or with a
second-order phase transition \cite{Bialas2,Hwa-int,Feynman}
depending on the values obtained. Thus an analysis of the
factorial moments may provide important information  on the
dynamical properties of the system. Ploszajczak and Tucholski were
the first to suggest searching for intermittency patterns in the
mass and charge distributions of the fragments produced in
energetic collisions \cite{Plo1}. Since then many studies show
that an intermittency pattern of fluctuations in the fragmentation
charge distributions has been observed in many data and models.
Much effort has been devoted to find the relation between
fragmentation, a possible critical behavior, and intermittency
\cite{Dorso,Phair,Elattari,Barz,Latora,Campi}.

Intermittency is defined by the relation
\begin{equation}
F_k(\delta') \equiv F_k(a\delta) = a^{-f(k)} F_k(\delta),
\end{equation}
between factorial moments $F_k(\delta') $ and $F_k(\delta_s)$
obtained for two different binning parameters $\delta$ and
$\delta'$ = a$\delta$.  Intermittency implies a linear relationship
in the double logarithmic plot of $ln F_k$ versus $-ln\delta$.

The fractal intermittency exponent, $f(k)$,  is related to the
factorial dimension $d_k$ by
\begin{equation}
f(k) = \frac{d_k}{k-1} > 0 .
\end{equation}
Different processes seem to give a different behavior of these
anomalous fractal dimension $d_k$: (1) $d_k$ = constant
corresponds to a monofractal, second order phase transition in the
Ising model and in the Feynman-Wilson fluid
\cite{Hwa-int,Feynman}. It has been also demonstrated that in the
case of a second order phase transition in the Ginzburg-Landau
description one gets $d_k = d_2 (k-1)^{\mu-1}$ with $\mu = 1.304$
\cite{Hwa-int}.  (2) $d_k \propto k$ correspond to multifractal,
cascading processes \cite{Bialas}. Therefore, a study of the
anomalous fractal dimensions can give useful information about the
evolution of the system.

Several models have been introduced 
to study the intermittency signal. One of the simplest models, widely used in
the analysis of experimental data and which gives intermittency,
is the percolation model. Percolation models predict a phase
transition corrected for finite size effects and produce, at the
critical point for this phase transition, a mass distribution
following a power law and obeying  scaling properties.

\begin{figure}
\begin{center}
\resizebox{.9\columnwidth}{!}{
\includegraphics{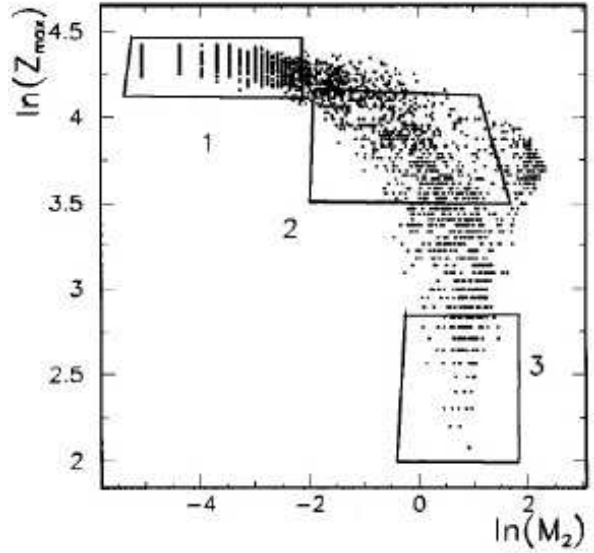}}
\end{center}
\caption{ Experimental Campi-scatter-plots from
Ref.~\cite{Mastinu_PRL}. Three cuts are
employed to selected the upper branch (1), the lower branch (3),
and the central region (2).} \label{INFN_Campi-plot}
\end{figure}

An intermittency analysis has been performed on many heavy ion
collision data as well as emulsion data. Here we give an example
of the multifragmentation data of Au + Au collisions at 35 MeV/u
which was performed at NSCL by the Multics-Miniball Collaboration
\cite{Mastinu_PRL}. A power-law charge distribution, $A^{-\tau}$
with $\tau \simeq  2.2$ and an intermittency signal has been
observed for the events selected in the region of the Campi
scatter-plot where "critical" behavior is expected. As shown in
Fig.~\ref{INFN_Campi-plot}, three cuts have been tested. The upper
branch is mostly related to the liquid branch and the lower branch
to the gas branch, while the central cut (2) is expected to belong
to a region where critical behavior takes place. Actually the
resultant charge distribution of  cut (2) shows a power-law
distribution with $\tau \simeq  2.2$ which is close to the droplet
model prediction if the liquid-gas critical point is explored. The
scaled factorial moments are shown in Fig. 14 for the different
cuts of Figure 13.  For cut 3, the logarithm of the scaled
factorial moment is always negative and almost independent of
-ln$\delta$; there is no intermittency signal. The situation is
different for cut 2 (the central part). The logarithm of the
scaled factorial moments is positive and almost linearly
increasing as a function of -ln$\delta $, and an intermittency has
been observed. Cut 1 gives a zero slope, no intermittency signal
again.

It has been argued that the interpretation of this experimentally
observed intermittency signal may, however, be problematic due to an
ensemble average effect \cite{Phair}. Since cut 2 involves  a
large range of impact parameters, the observed intermittency
signal could be an artifact of ensemble averaging, and can not be
seen as a definite evidence of large fluctuation driven by a
critical behavior.

\begin{figure}
\begin{center}
\resizebox{1.\columnwidth}{!}{
\includegraphics{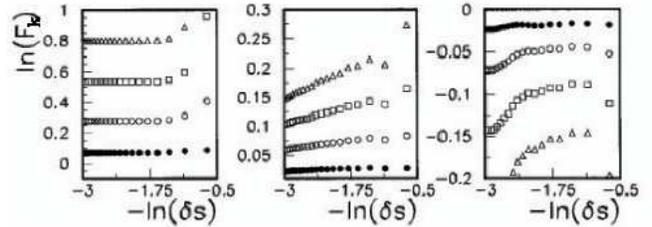}}
\end{center}
\caption{ Experimental results from
Ref.~\cite{Mastinu_PRL}. Scaled factorial moments ln($F_k$) vs
-ln($\delta_s$) for the three cuts made on
Fig.~\ref{INFN_Campi-plot}: left part cut 1, central part cut 2,
and right part cut 3. Solid circles represent the SFM of order k =
2, open circles k = 3, open squares k = 4, and open triangles k =
5. Figure is taken from Ref.~\cite{Mastinu_PRL}.}
\label{INFN_intm}
\end{figure}

Actually, several criticisms have been raised about the role of
the intermittency signal in nuclear fragmentation.
For instance, Elattari et al. showed that an
intermittency signal can be obtained even for a simple
fragmentation generator model by the random population of mass
bins with a power law distribution in which the only
non-statistical source of fluctuations is the mass conservation law
\cite{Elattari}. It has also been shown that the intermittency
signal is washed out when events of fixed total multiplicity are
selected \cite{Dorso,Campi} or when the size of the system tends
to infinity in the percolation model in which the fluctuations are
of nontrivial origin \cite{Campi}. Moreover, the intermittency
signal is not observed in the narrow excitation energy region
where the phase transition occurs in the framework of the
well-known Copenhagen statistical multi-fragmentation model
\cite{Barz} or in the data of 35-110 MeV/nucleon
$^{36}Ar+^{197}Au$ when the effects of impact parameter averaging
are reduced by some appropriate cuts \cite{Phair}.
However, it is important to notice that there is no reason to expect
intermittency if the phase transition is first order.

\begin{figure}
\begin{center}
\resizebox{.8\columnwidth}{!}{\vspace{-0.3truein}
\includegraphics{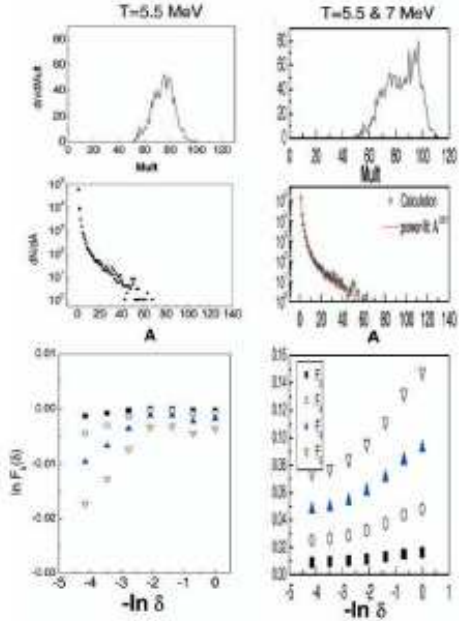}}
\end{center}
 \caption{Left panels: the multiplicity
 distribution (upper panel), the mass distribution (middle panel),
 the scaled factorial moments (bottom panel) with the multiplicity
restriction  for $^{129}Xe$ in the lattice gas model calculation.
Right panel: Same as the left panel but for the events mixed with
$T$ = 5.5 MeV and $T$ = 7 MeV.  Figure is taken from
Ref.~\cite{Ma_unpub}.} \label{intm_LGM}
\end{figure}

As an example, we check the intermittency behavior
\cite{Ma_unpub} in the Lattice Gas model for the disassembly  of
the system $^{129}Xe$ at 0.38$\rho_0$ in the framework of LGM (for
the details of the model description, please see the following
section). At a temperature $T$ = 5.5 MeV, the mass distribution
shows a power-law distribution with an effective power-law
parameter $\tau$ = 2.43. In a previous work with the same model,
it was shown that the liquid-gas phase transition
occurs near 5.5 MeV for this system in the LGM
\cite{Maprc99,Ma_PRL}. The $ln F_k$ shows slight negative values
with slightly positive slopes versus $-ln\delta$. However, this
kind of the positive slopes with a moment less than unity may be of
trivial origin and does not demonstrate the appearance of
intermittency which is characteristic of systems exhibiting
larger than Poisson fluctuations (i.e. the moment should be larger
than unity). In order to check the event mixture effect on the
scaled factorial moment, we mixed all the events at $T$ = 4 MeV
and $T$ = 7 MeV and also used the multiplicity cuts ($29\leq M
\leq 101$) and ($M < 29$ or $M > 101$) to see if   an
intermittency behavior can be found in such mixed events.
Figure~\ref{intm_LGM} shows these results. Even though all the
$ln F_k$ values are positive, they are  flat, i.e. there is no
intermittency signal. In these cases, the fluctuation is large
enough but the mass distribution shows no power-law distribution.
Hence, intermittency is absent. However, intermittency emerges
when the moments were calculated from the mixed events of $T$ =
5.5 MeV and $T$ = 7 MeV (Fig.~\ref{intm_LGM}). In this case,
the mass distribution shows a quite good power-law distribution
and fluctuations are also large enough to induce
intermittency.

From the above discussions, the apparent signals of
intermittency which emerge in many experimental data are not
easy to understand since many
experimental conditions bring some complexities to the pure signal
of intermittency, such as event mixing.
More precise experimental measurements
in the future are needed to probe the intermittency signal, which then may
be taken as a signal of true critical behavior.

 \section{Phenomenological Basis of Nuclear Zipf Law and Model Simulation}

In the above sections, we have focussed on the moment analysis, namely the
behavior of the
moments of the fragment size distribution, or of the scaled factorial
moments. Both are related to the fluctuations of some physical
observables. In this section, we would like to emphasize the
topological structure of the fragment size distribution, i.e. how
the fragments distribute from the largest to the smallest in
nuclear fragmentation. To this end, we introduce the Zipf-type
plot, i.e. rank-ordering plot, in the fragment size distribution as
well as Zipf's law which will be illustrated in the following
\cite{Ma_PRL,Ma_EPJ}.

 The original Zipf's law
\cite{Crystal} has been used for the diagnosis of nuclear
liquid-gas phase transition and as such we have called it the nuclear Zipf's
law. Zipf's law has been known as a statistical phenomenon
concerning the relation between English words and their frequency
in literature in the field of linguistics \cite{Crystal}. The law
states that, when we list the words in the order of decreasing
population, the frequency of a word is inversely proportional to
its rank \cite{Crystal}. This relation was found not only in
linguistics but also in other fields of sciences. For instance,
the law appeared in distributions of populations in cities,
distributions of income of corporations, distributions of areas of
lakes and cluster-size distribution in percolation processes
\cite{soc,Watanabe}. The details for the proposal of nuclear
Zipf's law can been found in Ref.~\cite{Ma_PRL,Ma_EPJ}. In this
report, we firstly define the nuclear Zipf plot for the fragment
mass (charge) distribution and nuclear Zipf's law in the
simulation with help of the lattice gas model. Then we show some
experimental  evidences for the nuclear Zipf law as well as some
remarks.

The tools we will use here are the isospin dependent lattice gas
model (LGM) and  molecular dynamical model (MD). The lattice gas
model was developed to describe the liquid-gas phase transition
for atomic systems by Lee and Yang \cite{Yang52}. The same model
has already been applied to nuclear physics for isospin
symmetrical systems in the grandcanonical ensemble \cite{Biro86}
with a sampling of the canonical ensemble
\cite{Maprc99,Camp97,Jpan96,Mull97,Jpan95,Jpan98,Gulm98}, and also
for isospin asymmetrical nuclear matter in the mean field
approximation \cite{Sray97}. In addition, a classical molecular
dynamical model is used to compare its results with the results of
the lattice gas model.

\begin{figure}
\begin{center}
\resizebox{0.80\columnwidth}{!}{%
  \includegraphics{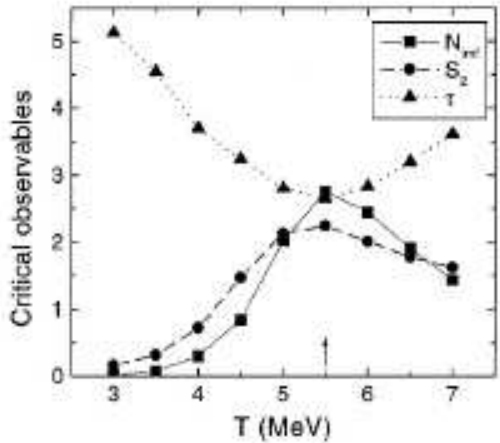}}

\end{center}
\caption{Effective power-law parameter, $\tau$,
second moment of the cluster distribution, $S_2$, and
multiplicity of intermediate mass fragments, $N_{imf}$ as a
function of temperature for the disassembly of $^{129}Xe$ at
$\rho_f \sim 0.38 \rho_0$ in I-LGM. The arrow represents the estimated temperature of
the phase transition. Figure is taken from \cite{Ma_EPJ}.}
\label{fig_obs}
\end{figure}

In the lattice gas  model, $A$ (= $N + Z$) nucleons with an
occupation number $s$ which is defined $s$ = 1 (-1) for a proton
(neutron) or $s$ = 0 for a vacancy, are placed on the $L$ sites of the
lattice. Nucleons in the nearest neighboring sites interact with
an energy $\epsilon_{s_i s_j}$. The hamiltonian is written as $E =
\sum_{i=1}^{A} \frac{P_i^2}{2m} - \sum_{i < j} \epsilon_{s_i
s_j}s_i s_j $. A three-dimension cubic lattice with $L$ sites is
used. The freeze-out density of disassembling system is assumed to
be
 $\rho_f$ = $\frac{A}{L} \rho_0$, where $\rho_0$ is the normal
 nuclear density. The disassembly of the system
is to be calculated at $\rho_f$, beyond which nucleons are too far
apart to interact.  Nucleons are put into lattice by Monte Carlo
Metropolis sampling. Once the nucleons have been placed we also
ascribe to each of them a momentum by Monte Carlo samplings of a
Maxwell-Boltzmann distribution. Once this is done the LGM
immediately gives the cluster distribution using the rule that two
nucleons are part of the same cluster if $P_r^2/2\mu -
\epsilon_{s_i s_j}s_i s_j < 0 $. This method is similar to the
Coniglio-Klein prescription \cite{Coni80} in condensed matter
physics and was shown to be valid in LGM
\cite{Camp97,Jpan96,Jpan95,Gulm98}. In addition, to calculate
clusters using MD we propagate the particles from the initial
configuration for a long time under the influence of the chosen
force. The form of the force is chosen to compare with the results
of LGM. The system evolves with the potential. At asymptotic times
the clusters are easily recognized. Observables based on the
cluster distribution in both models can now be compared. In
the case of proton-proton interactions, the Coulomb interaction
can also be added separately and it can be compared with the case
without Coulomb effects.

\begin{figure}
\begin{center}
\resizebox{0.85\columnwidth}{!}{%
  \includegraphics{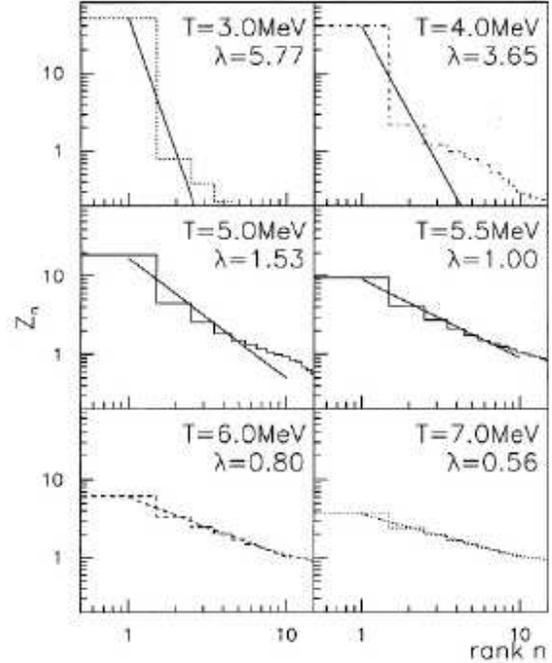}}
\end{center}
\caption{Average charge $Z_n$
with rank $n$ as a function of $n$ for $^{129}Xe$ $\rho_f \sim
0.38 \rho_0$ in I-LGM. The histograms are the calculation results and the
straight lines are their fits with $Z_n \propto n^{-\lambda}$.
Figure is taken from \cite{Ma_EPJ}.} \label{fig_obs2}
\end{figure}

In order to check the phase transition behavior in the
I-LGM, we will first show the calculations of some physical
observables in Fig.~\ref{fig_obs}, namely the effective power-law
parameter, $\tau$, the second moment of the cluster distribution,
$S_2$ \cite{Campi}, and the multiplicity of intermediate mass
fragments, $N_{imf}$, for the disassembly of $^{129}$Xe at the
freeze-out density $\rho_f \sim 0.38 \rho_0$ . These observables
have been successfully employed in previous works to probe the liquid-
gas phase transition, as shown in Refs.~\cite{Maprc99,Ma_EPJ,Jpan98}.
The valley of $\tau$, the peaks of
$N_{imf}$ and $S_2$ are located around $T$ $\sim$ 5.5 MeV which is the
signature of the onset of the phase transition.
Because of the exact mapping between the LGM and the Ising model,
we know that at this point the transition is first order.

\begin{figure}[h]
\begin{center}
\resizebox{0.75\columnwidth}{!}{%
  \includegraphics{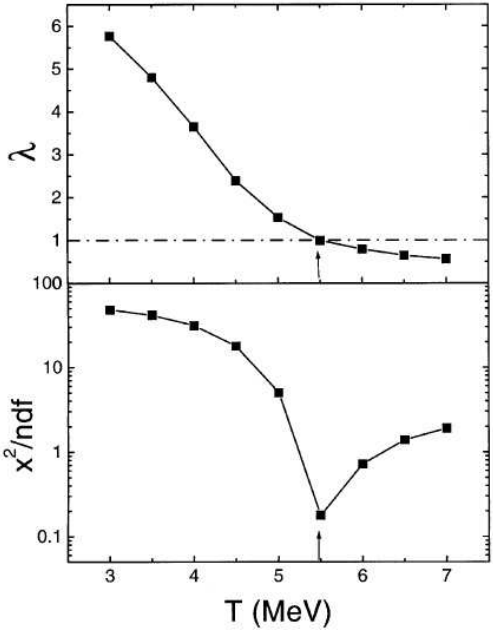}}
\end{center}
\caption{Slope parameter $\lambda$ of $Z_n$ to $n$ (top) and
$\chi^2$ test for Zipf's law (bottom) as a function of temperature
for $^{129}Xe$ at $\rho_f \sim 0.38 \rho_0$. The arrow represents
the estimated temperature of the phase transition. Figure is taken
from \cite{Ma_EPJ}.} \label{fig_obs3}
\end{figure}

Now we  present the results for testing Zipf's law in the charge
distribution of clusters. The law states that the relation between
the sizes and their ranks is described by $Z_n = c/n$ (n=1, 2, 3,
...), where $c$ is a constant and $Z_n$ (or $A_n$) is the average
charge (or mass) of rank $n$ in a charge (or mass) list when we
arrange the clusters in the order of decreasing size. For instance
the charge $Z_2$ of the second largest cluster with rank $n$ = 2
is one-half of the charge  $Z_1$ of the largest cluster, the
charge $Z_3$ of the third largest cluster with rank $n$ = 3 is
one-third of the charge $Z_1$ of the largest cluster, and so on.
In the simulations of this work, we averaged the charges for each
rank in charge lists of the events: we averaged the charges for
the largest clusters in each event, averaged them for the second
largest clusters, averaged them for the third largest clusters,
and so on. From the averaged charges, we examined the relation
between the charges $Z_n$ and their ranks $n$.
Figure~\ref{fig_obs2} shows such relations of $Z_n$ and $n$ for Xe
with different temperatures. The histogram is the simulated
results and the straight lines represent the fit with $Z_n \propto
n^{-\lambda}$ in the range of 1 $\leq$ $n$ $\leq$ 10, where
$\lambda$ is the slope parameter. $\lambda$ is 5.77 at $T$ = 3
MeV. Then we increased the temperature and examined the same
relation and obtained $\lambda$ = 3.65 and 1.53 at $T$ = 4 and 5
MeV, respectively. Up to $T$ = 5.5 MeV, $\lambda$ = 1.00, i.e., at
this temperature the relation is satisfied to the Zipf's law: $Z_n
\propto n^{-1} $. When the temperature increases, $\lambda$
decreases; for instance, $\lambda$ = 0.80 at $T$ = 6 MeV and
$\lambda$ = 0.56 at $T$ = 7. The temperature at which Zipf's law
emerges is consistent with the phase transition temperature
obtained in Fig.~\ref{fig_obs}, illustrating that the Zipf's law
is also an additional signal to determine the location of a phase
transition. From a statistical point of view, Zipf's law could
also be related to a critical phenomenon \cite{Fisher,Stauffer}.
The upper panel of Fig.~\ref{fig_obs3}  summarizes the parameter
$\lambda$ as a function of temperature.

In order to further illustrate that Zipf's law is most
probably fulfilled in phase transition points, we directly reproduce the
histograms with Zipf's law: $Z_n = c/n$. In this case, $c$ is the only
parameter, but what we are interested in is to check the
hypothesis of Zipf's law through a $\chi^2$ test. The bottom panel of
Fig.~\ref{fig_obs3}  shows the $\chi^2/ndf$ for the $Z_n$ -
$n$ relations at different $T$. As expected,  the minimum
$\chi^2/ndf$ is observed around the phase transition temperature, which
further indicates that Zipf's law of the fragment distribution
occurs around the liquid-gas phase transition point.

\section{Experimental Evidences of Nuclear Zipf Law}

\subsection{NIMROD results}

\begin{figure}
\begin{center}
 \resizebox{0.85\columnwidth}{!}{%
  \includegraphics{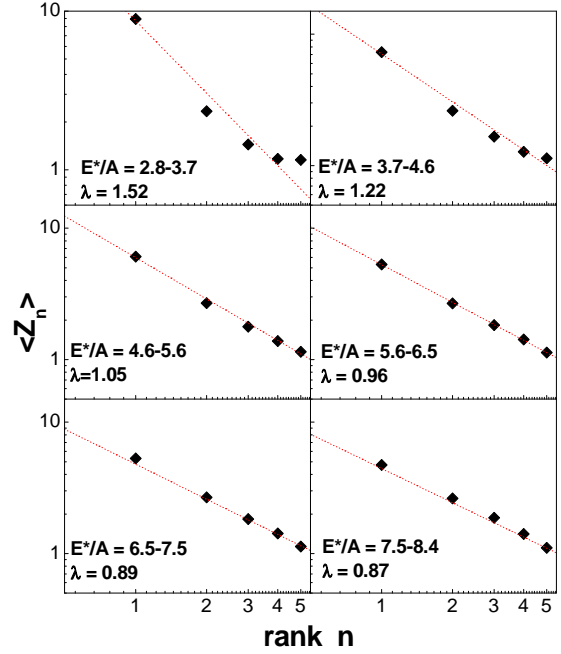}}

\end{center}
\caption{Zipf plots in six different excitation
energy bins for   the QP formed in $^{40}$Ar + $^{58}$Ni. The dots
are data and the lines are Zipf-law fits. The statistical error is
smaller than the size of the symbols.  } \label{fig_zipf}
\end{figure}

In Sec. 4.2, we gave some information on critical behaviors for the Texas A\&M
NIRMOD data based on the moment analysis technique .
Different signals of  critical behavior coherently pointing to the same excitation energy interval
have been shown. In this section, we will further show the
 significance of the 5-6 MeV region in NIMROD data
 using a Zipf's law analysis. In Fig.~\ref{fig_zipf} we
present Zipf plots for rank ordered average $Z$ in six different
energy bins. The lines in the figure are fits to the power law
expression $\langle Z_{n} \rangle \propto n^{-\lambda}$.
Figure~\ref{fluctuation}(f) shows the fitted Zipf exponent,
$\lambda$ parameter, as a function of excitation energy. As shown
in Fig.~\ref{fig_zipf}, this rank ordering of the observation
probability of fragments of a given atomic number, from the
largest to the smallest, does indeed lead to a Zipf's power law
parameter $\lambda$ = 1 in  the 5-6 MeV/nucleon range. Around this
excitation energy, the mean size of the second largest fragment is
1/2 of that of the largest fragment;  that of the third largest
fragment is 1/3 of the largest one, etc. This is a special kind of
size topology of fragment distributions, which is very different
from the equal-size fragment distribution expected if fragments
are formed through a spinodal instability inside the phase
coexistence region \cite{Borderie-WCI,Ber,spin1,spin2,Colo,Bord}.
This shows the relevance of using Zipf-plots to explore the
fragment size topology.

\subsection{CERN Emulsion Experiment}

The  nuclear Zipf-type plot has been also applied in the analysis
of CERN emulsion or Plastic data of Pb + Pb or Plastic at 158 AGeV
following Ma's proposal on Zipf law, and it was found that the
nuclear Zipf law is satisfied
in coincidence with other proposed signals of phase transition
\cite{Poland1,Poland2}.

Dabrowska et al. have extended these studies to the multifragmentation of
lead projectiles at an energy of 158 AGeV \cite{Poland1}. The
analyzed data were obtained from the CERN EMU13 experiment in
which emulsion chambers, composed of nuclear target foils and thin
emulsion plates interleaved with spacers, allow for precise
measurements of emission angles and charges of all projectile
fragments emitted from Pb-Nucleus interactions. The results on
fragment multiplicities, charge distributions and angular
correlations are analyzed for multifragmentation of the Pb
projectile after an interaction with heavy (Pb) and light (Plastic
- $C_5 H_4 O_2$) targets. A detailed description of the emulsion
experiment can be found in Ref.~\cite{Poland1}.

\begin{figure}
\begin{center}
\resizebox{0.85\columnwidth}{!}{%
  \includegraphics{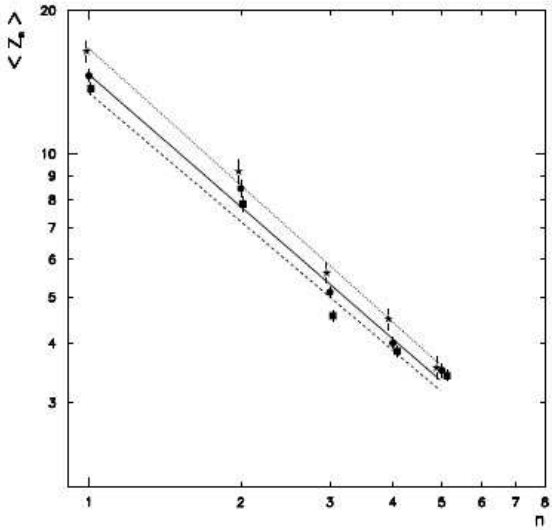}}
\end{center}
\caption{Zipf law fit to the dependences of the
mean charge of the fragment on its rank. The different symbols
represent the multifragmentation data of different beams with
an emulsion target. Circles and solid line represent Pb beam at 158
AGeV, squares and dashed line represent Au beam at 10.6 AGeV,
star and dotted line represent Au beam at 0.64 AGeV. Data are
taken from Ref.~\cite{Poland1}.} \label{fig_zipf_Poland1}
\end{figure}

Figure~\ref{fig_zipf_Poland1} shows the Zipf-type plot for
charged fragments heavier than helium emitted in
multifragmentation events of Au or Pb projectile at different
beam energies. The values of $\lambda$ exponents from fits
$\langle Z_n \rangle \sim n^{-\lambda}$ are 0.92 $\pm$ 0.03, 0.90
$\pm$ 0.02 and 0.96 $\pm$ 0.04 for beam energies of 158, 10.6
and 0.64 AGeV, respectively. Within the statistical errors, the
values of the $\lambda$ coefficient are the same in the studied
energy interval ( $<$ 1-158) AGeV and do not differ significantly
from unity \cite{Poland1}.

\begin{figure}
\begin{center}
\resizebox{0.90\columnwidth}{!}{%
  \includegraphics{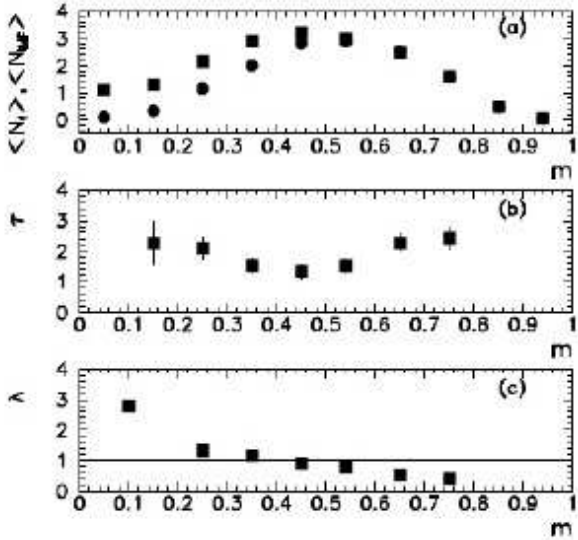}}
\end{center}
\caption{(a) Mean number, $\langle N_f \rangle$,
of fragments (squares) and mean number, $\langle N_{IMF} \rangle$,
of intermediate mass fragments (circles) as a function of the
normalized multiplicity m. Error bars are smaller than the size of
the squares and circles. (b) Power law exponent, $\tau$, of the
charge distribution of fragments in different intervals of m. (c)
Power law exponent, $\lambda$, in the Zipf's law (see text) in
different intervals of m. Error bars are smaller then data points.
The data is taken from Ref.~\cite{Poland2}.}
\label{fig_zipf_Poland2}
\end{figure}

Dabrowska et al. also studied the dependence of the power law
exponent $\lambda$ on the control parameter $m$, the normalized
multiplicity with respect to the total charge of spectator particles
\cite{Poland2}. In
 Fig.~\ref{fig_zipf_Poland2}(a) are shown the mean multiplicity
 $\langle N_f \rangle$ of fragments with $Z \geq 3$ and the mean number
 $\langle N_{IMF}\rangle$ of the intermediate fragments. The latter
 are usually defined as fragments with 3 $\leq Z \leq 30$.
In Fig.~\ref{fig_zipf_Poland2}(b) the dependence of the exponent
$\tau$ of the power fits to the charge distribution of fragments,
performed at different ranges of $m$, is also given. In this analysis, the fits
are restricted to fragment charges smaller than $Z = 16$. At small
values of $m$, a system has few light fragments and the power law
is steep; at large values of $m$ there are basically only many light fragments
leading again to a steep power law. At the moderate
excitation energies where heavier fragments appear and where we
expect the phase transition, the exponent $\tau$ has its lowest
value. As it can be seen from Fig.~\ref{fig_zipf_Poland2}(b), the minimum $\tau$
occurs for $m$ values between 0.35 and 0.55.
  In  Fig.~\ref{fig_zipf_Poland2}(c)  the dependence of $\lambda$
 obtained from the fits $\langle Z_n \rangle \sim n^{-\lambda}$,
 as a function of $m$ is depicted. The exponent $\lambda$ decreases with
 increasing $m$. Between $m\approx 0.3$ and $m\approx 0.5$ the value of $\lambda$ is
 close to unity and  Zipf's law is satisfied. This suggests that
 at this value of $m$ the liquid-gas phase transition might occur. It
 has been checked that $\lambda \sim 1$ occurs  in the same region of
 $m$ irrespectively of the mass of the target \cite{Poland2}. This means that the
 liquid-gas phase transition occurs when a given amount of energy
 is deposited into the nucleus and does not depend on the mass of
 the target. As expected, in the case of a liquid-gas phase transition, the previously shown maxima in
 frequency distributions of multiply charged fragments (Fig.~\ref{fig_zipf_Poland2}(a)
 ) as well as a minimum of the power law parameter $\tau$
(Fig.~\ref{fig_zipf_Poland2}(b)  ), all occur at the same values
of $m$, where Zipf's law emerges.

\subsection{Some Remarks on Zipf Law}

Campi et al. pointed out that for an infinite system, Zipf's law
is a mathematical consequence of a  power law cluster size
distribution with exponent $\tau \simeq 2$ \cite{Campi-Zipf}. More
precisely, both Zipf law exponent $\lambda$ and Fisher scaling
power-law exponent $\tau$ are connected through the formula
$\lambda=1/(\tau-1$) in {\it an} {\it infinite\ system} assuming
that the cluster size distribution is a power-law distribution.
They argued that such distributions appear at the critical point
with $\tau \simeq 2$ of many theories, eg. various theories of
cluster formation but also in the super-critical region of the
lattice-gas and realistic Lennard-Jones fluids \cite{Sator}.
However, the experimental fragment size distribution is mostly
neither power law distribution nor exponential distribution except
for some special situations. Also, the nuclear system is  always a
finite system, which means that the relationship between $\lambda$
and $\tau$ mentioned above is not strictly valid. To account for
finite size effects, Bauer et al. \cite{Bauer_Zipf} have evaluated
the fragment probabilities as a function of their rank at the
critical point for a finite system with fragment distributions
obeying to a finite size scaling ansatz. From this analytical
evaluation, where however the assumption is made that all
fragments including the  largest are much smaller than the source,
they suggest to extend the simple Zipf's law to a more general
Zipf-Mandelbrot distribution \cite{mandel1,mandel2}, $\langle
A_{r}\rangle = c (r + k)^{-\lambda}$, where the offset $k$ is an
additional constant that one has to introduce, and $\lambda$ is
asymptotically approximated as a function of the critical exponent
$\tau$, $\lambda = 1/(\tau-1)$ of the {\it infinite\ system}.

In any case, the Zipf-type plot is a
direct observable allowing to characterize the fragment hierarchy in
nuclear disassembly, and as such it is a useful signal of phase transition or critical behavior.

\section{Summary and Outlook}

In summary, the moment analysis method has been introduced and some
applications to nuclear multifragmentation have been
presented. Since we are dealing with a finite nucleus rather than
infinite nuclear matter, finite size effects must always be
discussed in the model calculations and
data analysis. Experimentally, the critical behavior of nuclear disassembly
can be investigated with the help of moment analysis. The occurrence of a fluctuation peak
 which can be extracted from the moment
analysis method can be interpreted as a signal of
critical behavior. Using the same analysis method as
for the percolation model, the liquid-gas universality class exponents are
approximately obtained in nuclear multifragmentation, such as in
EOS data and NIMROD data.
This would point to the observation of the
liquid-gas critical point or second order phase transition.
However, when we think about the system
size dependence of critical exponents and we consider some results using
Lattice Gas Model simulations and other related different analysis methods, it appears that some open
questions still remain concerning the order of the phase transition.
For instance, EOS Collaboration claimed
that there are continuous phase transitions for heavier systems,
namely Au and La and first order phase transitions for lighter
system, namely Kr. On the other hand, NIMROD data show critical behavior,
corresponding to a
continuous phase transition, for the light system Quasi-Ar.
Different conclusions are then reached for similar
light systems.
Recent systematic analyses of
caloric curves \cite{JBN0,Ma2005,JBN-WCI} and configurational
energy fluctuations \cite{Gul-WCI}, indicate that heavier
systems may  undergo a first order phase transition while
lighter systems can probably sustain a higher temperature,
possibly even above the critical point, which would make the first
order phase transition observed in heavy nuclei become a cross-over
in lighter systems. Concerning configurational energy fluctuations, a well
pronounced peak at an excitation energy around 5 MeV was shown in
Multics, Indra, Isis and Nimrod data \cite{Gul-WCI}. However, this
fluctuation appears monotonically decreasing in EOS data
\cite{SrivastavaPRC65}. Thus it deserves further
investigations.

Scaled factorial moments and intermittency have also been reviewed and
some examples given to show the apparent intermittency in
nuclear fragmentation. However, some complex ingredients in
experimental measurements, such as mixtures of event
multiplicities or temperature fluctuations in the data can induce
spurious intermittency-like behavior which implies that the apparent
"intermittency" can not be taken as a unique signal of the
critical behavior. Without 4$\pi$ detector upgrades allowing better data sorting,
it remains difficult to take apparent "intermittency" behaviors as a
signal of critical behavior in nuclear multifragmentation.

Finally, nuclear Zipf-type plots are introduced and \linebreak[4]
Zipf's law is proposed to be related to a phase transition or a
critical behavior of nuclei. Around the transition point, the
cluster mass (charge) shows inversely to its rank, i.e. Zipf's law
appears. Even though the criterion is phenomenological, it is a
simple and practicable tool to characterize the fragment hierarchy
in nuclear disassembly. The 4$\pi$ multifragmentation data of
heavy ion collision at Texas A$\&$M University and  the CERN
emulsion/plastic data exhibit the Zipf law around the same
excitation energy deposit.  The satisfaction of the Zipf law for
the cluster distributions illustrates that the clusters obey at
this point a particular rank ordering distribution  very different
from the equal-size fragment distribution which may occur due to
spinodal instability inside the liquid-gas coexistence region. To
conclude, we should mention that all these transition signals,
such as the fluctuation peak, critical exponents, Fisher scaling
as well as Zipf's law etc may not be very robust individually
since we are facing a transient finite charged nuclear system. A
unique signal can not give any definite information as to whether
the system is in a critical point or is undergoing a phase
transition. Only many coherent signals emerging together can
corroborate the observation of a phase transition or a critical
behavior in finite nuclei.

This work was supported in part by the Shanghai Development
Foundation for Science and Technology under Grant Number
05XD14021, the National Natural Science Foundation of China  under
Grant Nos. 19725521, 10328259, 10135030, 10535010 and the Major
State Basic Research Development Program under Contract No
G200077404.

%

\end{document}